\documentclass[pdflatex,sn-mathphys-num]{sn-jnl}

\usepackage{graphicx}%
\usepackage{multirow}%
\usepackage{amsmath,amssymb,amsfonts}%
\usepackage{amsthm}%
\usepackage{mathrsfs}%
\usepackage[title]{appendix}%
\usepackage{xcolor}%
\usepackage{textcomp}%
\usepackage{manyfoot}%
\usepackage{booktabs}%
\usepackage{algorithm}%
\usepackage{algorithmicx}%
\usepackage{algpseudocode}%
\usepackage{listings}%
\usepackage{tcolorbox}

\theoremstyle{thmstyleone}%
\theoremstyle{thmstyletwo}%

\theoremstyle{thmstylethree}%

\begin{document}

\title[Article Title]{Single Shot AI-assisted quantification of KI-67 proliferation index in breast cancer}


%
%
%
%
%
%

\author[1]{\fnm{Deepti} \sur{Madurai Muthu}}

\author[2]{\fnm{Priyanka} \sur{S}}

\author[1]{\fnm{Lalitha Rani} \sur{N}}

\author*[3]{\fnm{P. G.} \sur{Kubendran Amos}}\email{prince@nitt.edu}

\affil[1]{\orgdiv{Department of Pathology}, \orgname{KAP Viswanatham Government Medical College}, \orgaddress{\city{Tiruchirappalli}, \postcode{620001}, \state{Tamil Nadu}, \country{India}}}

\affil[2]{\orgdiv{SCOPE}, \orgname{Vellore Institute of Technology-AP}, \orgaddress{\city{Amaravati}, \postcode{522237}, \state{Andhra Pradesh}, \country{India}}}

\affil*[3]{\orgdiv{Theoretical Metallurgical Group, Department of Metallurgical and Materials Engineering}, \orgname{National Institute of Technology}, \orgaddress{\city{Tiruchirappalli}, \postcode{620015}, \state{Tamil Nadu}, \country{India}}}


\abstract{Reliable quantification of Ki-67, a key proliferation marker in breast cancer, is essential for molecular subtyping and informed treatment planning. Conventional approaches, including visual estimation and manual counting, suffer from interobserver variability and limited reproducibility. This study introduces an AI-assisted method using the YOLOv8 object detection framework for automated Ki-67 scoring. High-resolution digital images (40x magnification) of immunohistochemically stained tumor sections were captured from Ki-67 hotspot regions and manually annotated by a domain expert to distinguish Ki-67-positive and negative tumor cells. The dataset was augmented and divided into training (80$\%$), validation
(10$\%$), and testing (10$\%$) subsets. Among the YOLOv8 variants tested, the Medium model achieved the highest performance, with a mean Average Precision at 50$\%$ Intersection over Union (mAP50) exceeding 85$\%$ for Ki-67-positive cells. The proposed approach offers an efficient, scalable, and objective alternative to conventional scoring methods, supporting greater consistency in Ki-67 evaluation. Future directions include developing user-friendly clinical interfaces and expanding to multi-institutional datasets to enhance generalizability and facilitate broader adoption in diagnostic practice.}




\maketitle

\section{Introduction}\label{sec1}

Breast cancer (BC) is the most commonly diagnosed cancer among women worldwide and the second leading cause of cancer-related deaths in females [Sung H et al., 2021]. Advances in molecular classification have transformed breast cancer management, enabling more precise treatment approaches. BC is classified into four primary molecular subtypes-Luminal A, Luminal B, HER2-enriched, and Triple-Negative (basal-like) breast cancer-based on the status of hormone receptors (Estrogen Receptor [ER] and Progesterone Receptor [PR]) and human epidermal growth factor receptor 2 (HER2) [Blows FM et al., 2010]. This classification is essential, as different subtypes exhibit distinct biological behaviors, prognoses, and responses to therapy.

Among these subtypes, Luminal A and Luminal B tumors are hormone receptor-positive but differ significantly in their proliferation rates and therapeutic strategies. Luminal A tumors are ER and/or PR positive, HER2 negative, and exhibit a low Ki67 index, reflecting a slower proliferation rate and a more favorable prognosis. These tumors generally respond well to endocrine therapy alone. In contrast, Luminal B tumors are ER and/or PR positive, HER2 positive or negative, with a high Ki67 index, indicating a more aggressive nature and the need for additional chemotherapy alongside endocrine therapy [Soliman NA, 2016]. Thus, accurately distinguishing between these subtypes is critical in optimizing treatment decisions.

Immunohistochemistry (IHC) is a key technique in pathology for detecting biomarker expression in tissue samples. In breast cancer, Ki67 IHC specifically quantifies tumor proliferation, aiding in prognosis and treatment selection. Ki67 IHC plays a key role in breast cancer prognostication. Ki67 is a nuclear non-histone protein expressed in all active phases of the cell cycle except G0, making it a reliable indicator of tumor cell proliferation. Its expression reflects the growth fraction of a tumor, aiding in prognostic assessment. It serves as a dynamic biomarker for assessing tumor proliferation [Nishimura R et al., 2010]. The Oncotype DX assay, which includes Ki67 among its 21 genes, highlights its role in predicting recurrence risk and guiding adjuvant therapy selection [Paik et al., 2004].

The clinical importance of Ki67 in therapeutic decision-making was further reinforced by the 2021 St. Gallen/Vienna Consensus Conference and the International Ki67 in Breast Cancer Working Group, which recommended avoiding adjuvant chemotherapy for patients with Ki67
$<$5$\%$ while strongly advising it for those with Ki67 >30$\%$ [Thomssen C et al., 2021]. These
guidelines emphasis the need for accurate and reproducible Ki67 quantification, as even small variations in assessment can impact treatment choices.

Despite its clinical significance, Ki67 scoring remains a subject of debate due to the lack of standardized evaluation methods [Petrelli F et al., 2015]. In 2011, the International Ki67 in Breast Cancer Working Group highlighted the inconsistencies in Ki67 assessment, emphasizing the urgent need for standardization in clinical practice [Dowsett M et al., 2011]. Traditionally, pathologists rely on visual assessment under a microscope, which is prone to interobserver variability and limited reproducibility. Manual cell counting, requiring the analysis of at least 500-1000 tumor cells to achieve an acceptable error rate, is both time- consuming and susceptible to subjective biases [Polley M et al., 2013; Gudlaugsson E et al., 2010]. Given these limitations, the integration of artificial intelligence (AI) in pathology offers a promising solution for improving the accuracy and efficiency of Ki67 assessment.

AI-driven image analysis can standardize Ki67 quantification by automating cell detection, improving pathologist performance in Ki67 scoring [Nature, 2024], thereby reducing observer variability and enhancing reproducibility. Deep learning-based models can rapidly analyze large datasets, effectively evaluating Ki67 digital image analysis in invasive breast carcinoma [Wiley, 2024], identify Ki67-positive and negative cells with high precision, and generate objective, quantifiable scores. This has significant implications for pathologists, as AI-assisted systems can serve as decision-support tools, optimizing workflow efficiency and ensuring consistent evaluations.

By leveraging AI for Ki67 IHC analysis, this study aims to bridge the gap between computational advancements and clinical applicability, ultimately contributing to more reliable and reproducible breast cancer prognostication.. 

\section{DATA PROCESSING AND TECHNIQUE}\label{sec:theory}

\begin{figure}[h]
    \centering
    \includegraphics[width=\linewidth]{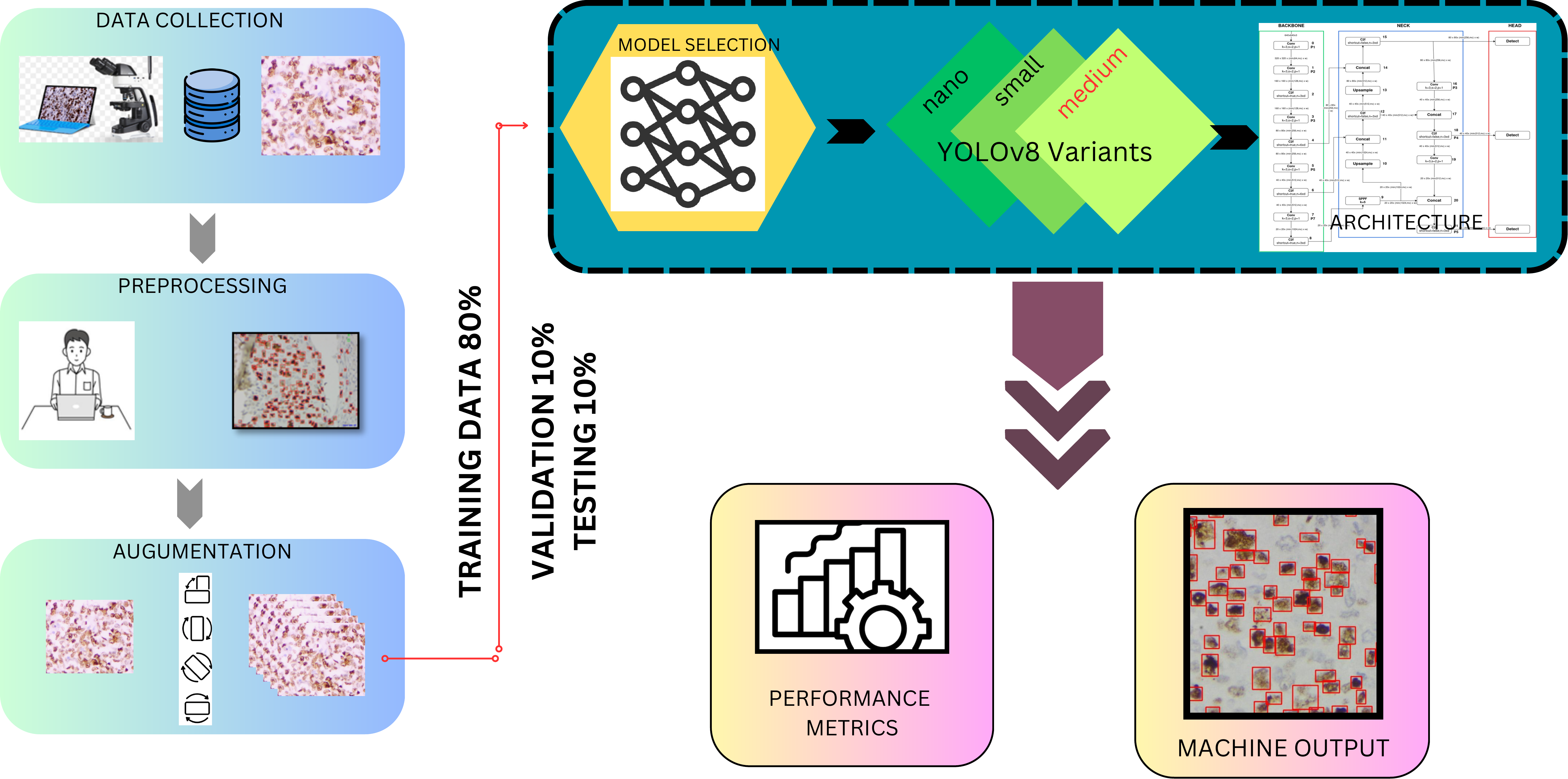} 
    \caption{The Ki67 Detection Workflow using YOLOv8 begins with the data collection of 180 Ki67-labeled IHC images, which are manually annotated by domain experts to label Ki67-positive (red) and Ki67-negative (green) tumor cells. During preprocessing, images are resized to 640 x 640 pixels, normalized, and converted into an appropriate color space to ensure consistency in model training. To enhance dataset diversity, augmentation techniques such as rotation, flipping, scaling, and noise addition are applied, expanding the dataset to 1,863 images and improving model robustness. For model selection, three YOLOv8 variants - Nano, Small, and Medium - are evaluated, with 80$\%$ of the dataset allocated for training and 10$\%$ for validation to optimize model parameters and prevent overfitting. The model's performance is assessed using precision, recall, and mAP50, ensuring reliable detection of Ki67-positive and negative tumor cells. After evaluation, YOLOv8 Medium is identified as the best-performing model, demonstrating superior bounding box regression and detection accuracy. The final model is tested on a separate testing subset, confirming its generalization capability. In the clinical integration phase, the validated model is deployed into pathology workflows for automated Ki67 scoring, reducing manual variability and improving diagnostic consistency. The results are provided with bounding box overlays, allowing pathologists to efficiently assess tumor proliferation, aiding in prognosis and treatment planning.}
    \label{fig:wrokflow}
\end{figure}

\subsection{Tissue Processing and Image Acquisition}

Formalin-fixed, paraffin-embedded (FFPE) tissue blocks from 30 diagnosed cases of invasive ductal carcinoma were retrieved from the archives of the Department of Pathology, KAP Viswanatham Government Medical College, Tiruchirappalli. Approval from the Institutional Ethics Committee was obtained to ensure ethical standards.

Hematoxylin and eosin (H$\&$E) staining was performed to select representative tumor sections for analysis. Subsequently, immunohistochemical staining was done using the MIB-1 clone of the Ki-67 antibody, a nuclear protein associated with cellular proliferation. This staining resulted in brown nuclear signals indicating Ki-67 positivity against a blue hematoxylin counterstain background, highlighting proliferating tumor cells.

In IHC-stained slides of Ki-67, hotspot regions are areas with the highest density of Ki-67- positive tumor cells. These regions are crucial because Ki-67 expression is often heterogeneous within a tumor, and selecting the most proliferative regions provides a more reliable measure of tumor aggressiveness. These were identified by an experienced pathologist in accordance with the International Breast Cancer Ki-67 Working Group's recommendations, which suggest evaluating at least 500 cells in the most proliferative areas to ensure consistency and reproducibility in Ki-67 scoring.

High-resolution digital images of these hotspot regions were acquired at 40x magnification using a Magvision camera integrated with the microscope's optical system. For each case, 6 to 7 images were captured to ensure adequate representation of tumor heterogeneity. The camera was connected to a computer workstation for optimized image acquisition and storage. A total of 180 images were systematically obtained for analysis.

\subsection{Annotation and Augumentation}

In our study, we employed a comprehensive approach to data annotation and augmentation to enhance the performance of our AI model in detecting Ki-67-positive tumor cells. Initially, raw images were processed using Roboflow, a platform that streamlines image preprocessing and augmentation. Images were resized to 640 x 640 pixels while maintaining the aspect ratio, with necessary padding added. Pixel values were normalized to a range of 0 to 1 to standardize the input data. Processed images were converted into tensors (B x C x H x W format) using the PyTorch framework, enabling efficient data handling and batch processing through GPU acceleration. Roboflow's auto-orientation feature corrected any misalignments, and grayscale conversion was applied where appropriate to enhance contrast and feature detection.

Manual annotation was performed using the "LabelMe" Python tool, which is known for its efficiency in creating polygonal annotations. Bounding boxes were drawn to label two classes: Ki67-positive tumor cells and Ki67-negative tumor cells. Given the small size of tumor cells, the Intersection over Union (IoU) threshold was optimized to 0.5, ensuring accurate differentiation between overlapping tumor cells. This manual annotation, supervised by an expert pathologist, established a ground truth dataset for training the AI algorithm.

To enhance the robustness and diversity of the dataset, various augmentation techniques were employed. These included horizontal and vertical flips, 90$^0$ rotations (clockwise and counterclockwise), 180$^0$ rotations, random cropping ranging from 0$\%$ to 8$\%$ to simulate variability in tissue sectioning, and brightness adjustments between -24$\%$ to +24$\%$ to account for staining intensity variations. These augmentation strategies expanded the dataset to a total of 1,863 images, enhancing the model's ability to generalize across diverse image conditions.

During the object detection phase, Non-Maximum Suppression (NMS) was applied to refine object detection by eliminating redundant bounding boxes, thereby improving the precision of
the model in identifying Ki67-positive and negative cells. In pathology images, where dense cell clusters often lead to overlapping detections, NMS was employed with a threshold of 0.3 to suppress redundant bounding boxes while preserving accurate Ki-67-positive cell identification.

This structured approach to annotation and augmentation was crucial in developing a robust and accurate AI model for analyzing Ki-67 immunohistochemical staining in breast cancer pathology.

\subsection{Object Detection Technique}

\subsubsection{Selection of YOLOv8 Model}

In this study, we selected the YOLOv8 model for its superior performance in real-time object detection tasks. Compared to conventional convolutional neural network (CNN) classifiers like ResNet, DenseNet, and EfficientNet, YOLOv8 offers faster detection speeds and improved localization capabilities, which are critical for accurately identifying Ki67-positive tumor cells in histopathological images. Pathological image analysis presents unique challenges such as high-resolution images, densely packed cells, color variability, and variations in staining intensity. Traditional CNN-based classifiers, primarily designed for classification tasks, face difficulties in detecting small, spatially clustered objects due to their hierarchical feature extraction methods. These models process entire images in a sliding window manner or depend on region proposal networks (RPNs) for object detection, increasing computational costs and inference times.

YOLOv8 addresses these limitations by employing a one-stage detection framework, which processes the image in a single pass, enabling real-time detection while maintaining high accuracy. Unlike earlier CNNs and YOLO versions, YOLOv8 introduces an anchor-free approach to bounding box prediction, eliminating the dependency on predefined anchor boxes. This innovation reduces computational complexity and enhances model efficiency, particularly in detecting small and overlapping tumor cells by dynamically adjusting bounding boxes based on object characteristics. Additionally, YOLOv8 integrates an advanced C2f module (Cross- Stage Partial Networks with Fusion), which enhances feature representation, allowing for finer localization of tumor regions. The model's decoupled head for classification and regression improves detection precision by separating classification and bounding box regression, thereby minimizing misclassification errors between Ki67-positive and Ki67-negative tumor cells.

Advancements Over Predecessors: YOLOv8 introduces several enhancements over its predecessors, such as improved feature extraction and localization accuracy, making it well- suited for medical image analysis. The model?s architecture includes mosaic data augmentation, anchor-free detection, a C2f module, a decoupled head, and a modified loss function, all contributing to its superior performance. Additionally, YOLOv8 integrates a Feature Pyramid Network (FPN), which allows multi-scale learning, enabling the model to detect Ki-67-positive nuclei of varying sizes effectively. This is particularly beneficial in
pathology images where differences in magnification levels and tumor heterogeneity affect object scale. Compared to models like PathoNet and MobileUnet, which rely on segmentation- based approaches requiring pixel-wise classification, YOLOv8 directly detects whole Ki-67- positive tumor cells in one-shot object detection, significantly reducing processing time while maintaining detection accuracy. Furthermore, YOLOv8 incorporates Bidirectional Feature Pyramid Network (BiFPN) multi-scale feature fusion, which enhances performance across varying magnifications and image resolutions.

The use of Non-Maximum Suppression (NMS) with optimized thresholds and advanced filtering techniques further refines object detection by reducing redundant bounding boxes, improving precision in distinguishing Ki-67-positive cells from staining artifacts. Additionally, false positives were reduced through enhanced confidence thresholding, dynamic receptive field adaptation, and adaptive feature learning, which helped the model differentiate between true Ki67-expressing nuclei and background staining artifacts. The dynamic receptive field mechanism further improves localization by allowing the model to focus on relevant cellular structures while ignoring noise.

\subsubsection{Model Variants and Training Strategy}

\begin{figure}[h]
    \centering
    \includegraphics[width=\linewidth]{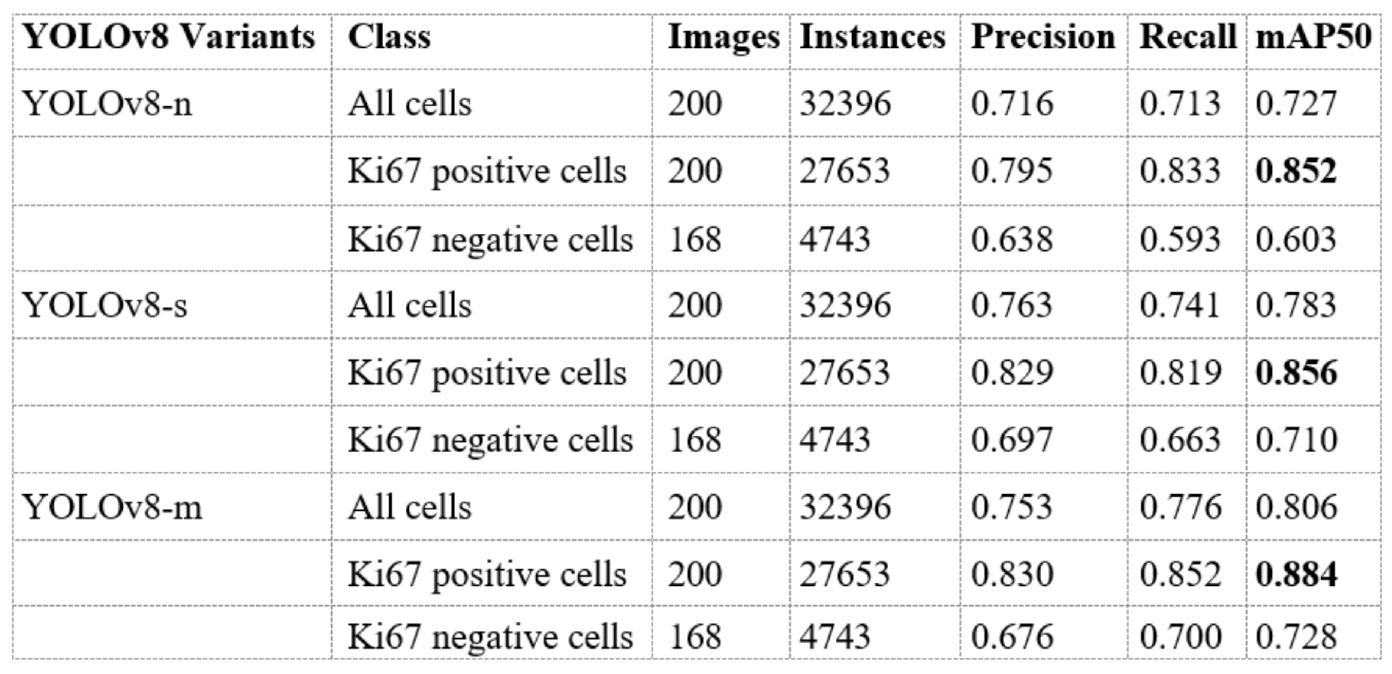} 
    \caption{Comparative Performance of YOLOv8 Variants for Ki67 Cell Detection}
    \label{fig:table}
\end{figure}

Variants Utilized: To balance detection accuracy and computational efficiency, we employed three YOLOv8 variants:
\begin{enumerate}
 \item Nano: Optimized for speed with minimal computational requirements.
 \item Small: Offers a balance between speed and accuracy.
 \item Medium: Prioritizes accuracy, suitable for detailed analysis.
\end{enumerate}

Dataset Splitting: The dataset was divided into training 1556 images (80$\%$), validation 200 images (10$\%$), and testing 107 images (10$\%$) subsets to ensure robust model evaluation.

\subsubsection{Fine-Tuning with Pretrained Models}

Transfer Learning Approach: Pretrained models from the COCO dataset, initially trained on general object detection tasks, were fine-tuned on our annotated histopathology dataset for domain-specific adaptation. This transfer learning approach leverages existing knowledge to improve performance in identifying Ki67-positive tumor cells. By implementing YOLOv8 and its variants, we aimed to achieve accurate and efficient detection of Ki67-positive tumor cells in histopathological images, thereby contributing to advancements in medical image analysis.

\section{Results and Discussion}

\subsection{Proficiency Parameters and Their Relevance}

Precision is the ratio of correctly predicted positive cases to the total predicted positive cases, represented as , where TP is True Positive and FP is False Positive. Precision ensures accurate identification of Ki67-positive and negative tumor cells, crucial for precise clinical decision-making in this study.

Recall is the ratio of correctly identified positive cases to the total actual positive cases, defined as , where FN represents False Negative. High recall reduces missed detections, particularly important for Ki67-positive cells, which directly influence treatment strategies.

\begin{figure}[h]
    \centering
    \includegraphics[width=\linewidth]{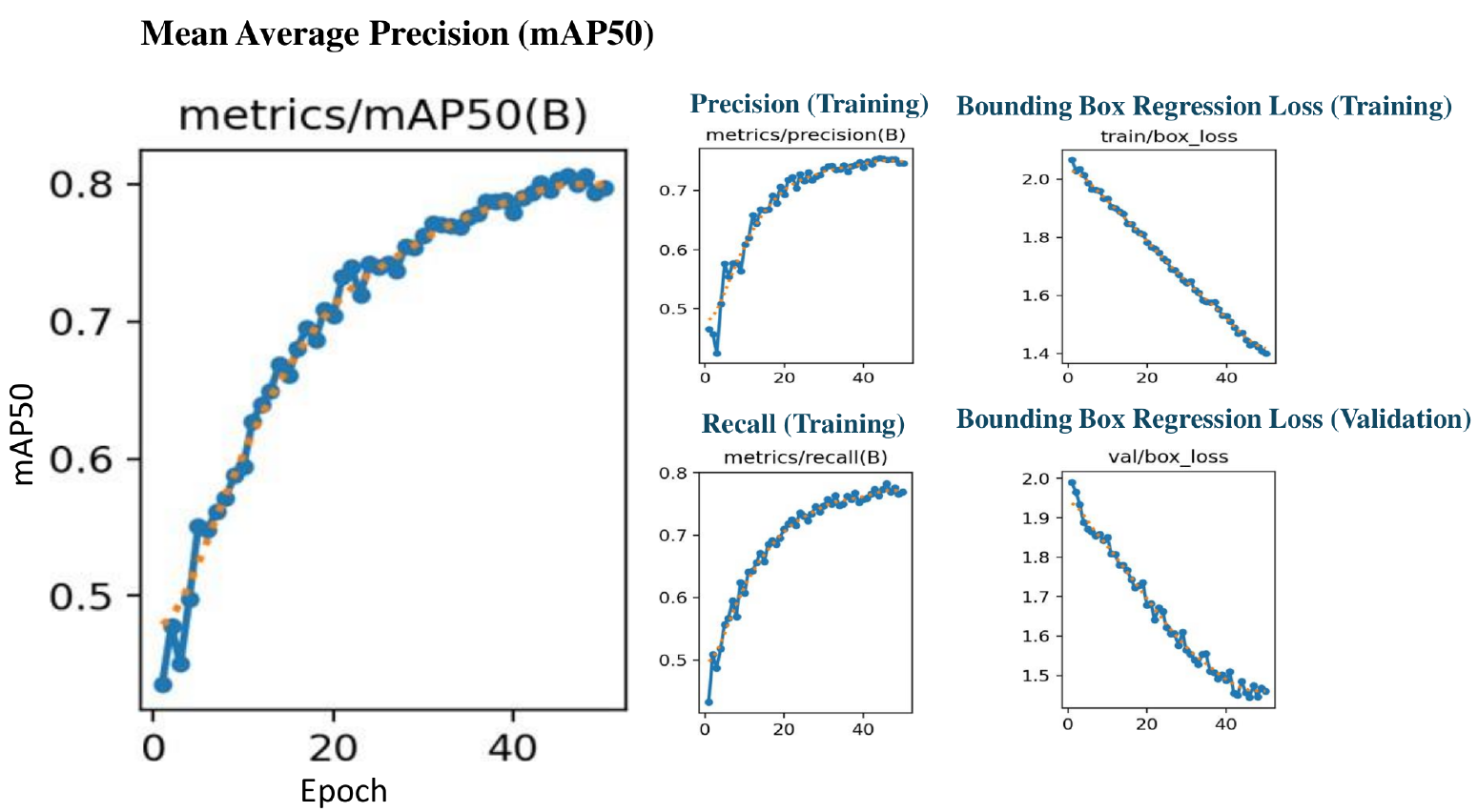} 
    \caption{Increase in the performance of the best YOLOv8 variant.}
    \label{fig:graphpdf}
\end{figure}

Mean Average Precision (mAP50) is the average precision calculated at an Intersection over Union (IoU) threshold of 0.5, reflecting overall detection accuracy and localization quality (Everingham et al., 2010). This metric integrates precision and recall, providing a comprehensive assessment of model performance.

The performance variation observed in YOLOv8 variants-Nano, Small, and Medium-is attributable to differences in their neural network depth and complexity. The Nano variant is optimized for minimal computational resources, offering speed at the expense of accuracy.
The
Small variant balances computational efficiency and detection accuracy, whereas the Medium variant, with a deeper backbone and enhanced feature extraction, demonstrates superior precision and recall, resulting in the highest mAP50 scores.

The higher mAP50 achieved for Ki67-positive cells ($>$85$\%$) compared to Ki67-negative cells (73$\%$) is primarily due to the significantly higher number of annotated Ki67-positive cell instances (27,653) compared to Ki67-negative cell instances (4,743). This dataset imbalance enhances training efficiency, resulting in improved detection accuracy for Ki67-positive cells. Additionally, staining quality plays a critical role; Ki67-positive cells typically exhibit pronounced dark brown-to-maroon staining, providing clear visual cues that facilitate accurate detection. Conversely, Ki67-negative cells often display weak or inconsistent staining intensity, sometimes merging visually with the background, complicating their accurate detection.

\subsection{YOLOv8 Medium as the Best Model}

In evaluating the YOLOv8 variants-Nano, Small, and Medium-we systematically assessed performance metrics across 50 training epochs to identify the most suitable model for accurately detecting Ki67-positive and negative tumor cells.

The consistent decrease in training and validation Bounding Box Regression Loss indicates the model's improving accuracy in aligning predicted bounding boxes closely with ground-truth annotations, a critical aspect of precise tumor cell localization (Redmon et al., 2016). YOLOv8 Medium exhibited significant improvements compared to Nano and Small variants, demonstrating superior localization capabilities.
Precision and recall values during training for all variants revealed improving trends, stabilizing at high values in later epochs. These metrics indicate that the models consistently reduced false positives (precision) and minimized missed detections (recall), thereby reliably aligning predicted bounding boxes with annotated ground truth. Improved precision and recall directly enhance clinical reliability by reducing diagnostic errors.

A critical distinguishing factor emerged upon examining the mean Average Precision (mAP50), which effectively integrates both precision and recall at an Intersection over Union (IoU) threshold of 0.5. While all variants displayed improving trends, the YOLOv8 Medium variant achieved the highest and most stable mAP50 ($>$85$\%$), clearly surpassing the Nano and Small variants. This superior mAP50 score underscores YOLOv8 Medium's optimized balance between precision and recall, attributed to its deeper neural network architecture and enhanced feature extraction capabilities, resulting in fewer false-positive and false-negative detections.

\begin{figure}[h]
    \centering
    \includegraphics[width=\linewidth]{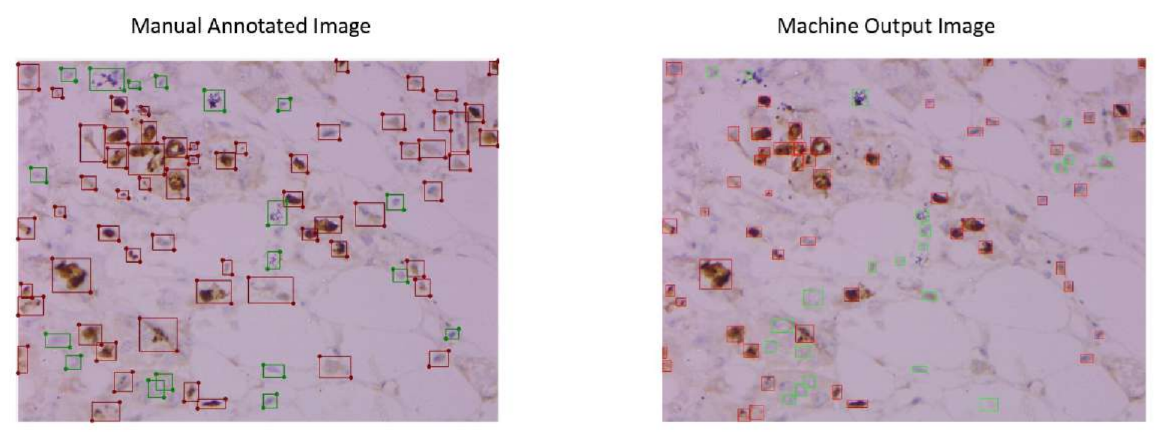} 
    \caption{Representative images demonstrating the YOLOv8 Medium variant performance in detecting Ki67-positive and negative tumor cells. Red bounding boxes indicate accurate detection of Ki67-positive tumor cells, closely matching ground-truth annotations with minimal false positives. Green bounding boxes represent Ki67-negative tumor cells, showing occasional missed detections or minor localization discrepancies, leading to false negatives, consistent with the quantitative analysis. These visual outputs confirm model proficiency, highlighting clinical reliability for precise Ki67 proliferation index assessment.}
    \label{fig:image_jpg}
\end{figure}

Thus, based on comprehensive analysis across bounding box regression accuracy, precision, recall, and overall detection performance (mAP50), YOLOv8 Medium distinctly emerges as the most proficient model for accurately and consistently determining the Ki67 proliferation index in clinical settings. Enhanced precision and recall values directly improve clinical reliability by reducing diagnostic errors.

\section{Conclusion}

Ki67 is a key biomarker in breast cancer prognostication, aiding in tumor classification and treatment decisions. However, manual assessment is subjective and prone to variability. This study overcomes these challenges by leveraging the YOLOv8 Medium variant to automate and standardize Ki67 scoring, enhancing diagnostic consistency and reliability.

Among the evaluated models, the YOLOv8 Medium variant emerged as the optimal choice, demonstrating superior performance in identifying Ki67-positive and negative tumor cells. This model achieved a high mAP50 score ($>$85$\%$) due to its deeper neural architecture, efficient bounding box regression, and robust feature extraction capabilities, ensuring precise localization and classification of tumor cells.

Notably, despite the moderate dataset size, the YOLOv8 Medium variant achieved high detection accuracy (mAP50 $>$85$\%$), highlighting its capability to effectively generalize even with limited training data-a crucial advantage in medical image analyses, where extensive annotated datasets can be challenging to obtain (Sun et al., 2017). This finding underscores the model's ability to provide reliable results even in data-limited scenarios.

A key limitation of the current approach is its accessibility. At present, model implementation requires computational expertise, making it inaccessible to pathologists without programming knowledge. Bridging this gap requires the development of user-friendly interfaces that enable direct AI interaction without coding expertise, allowing seamless clinical integration. Additionally, expanding the dataset to include more diverse, multi-institutional samples will enhance model generalizability and robustness. Ultimately, integrating this AI-driven system into clinical workflows with an intuitive interface will enable pathologists to directly utilize automated Ki67 scoring, ensuring broader accessibility and clinical adoption.


\section{References}

\begin{enumerate}
    \item Sung H et al., 2021 - Sung, H., Ferlay, J., Siegel, R. L., Laversanne, M., Soerjomataram, I., Jemal, A., \& Bray, F. (2021). Global Cancer Statistics 2020: GLOBOCAN Estimates of Incidence and Mortality Worldwide for 36 Cancers in 185 Countries. \textit{CA: A Cancer Journal for Clinicians}, 71(3), 209-249.
    
    \item Blows FM et al., 2010 - Blows, F. M., Driver, K. E., Schmidt, M. K., Broeks, A., Van Leeuwen, F. E., Wesseling, J., ... \& Pharoah, P. D. P. (2010). Subtyping of breast cancer by immunohistochemistry to investigate a relationship between subtype and short- and long-term survival: a collaborative analysis of data for 10,159 cases from 12 studies. \textit{PLoS Medicine}, 7(5), e1000279.
    
    \item Soliman NA, 2016 - Soliman, N. A., \& Yussif, S. M. (2016). Ki-67 as a prognostic marker according to breast cancer molecular subtype. \textit{Cancer Biology \& Medicine}, 13(4), 496-504.
    
    \item Nishimura R et al., 2010 - Nishimura, R., Takahashi, T., Iwamoto, T., Niikura, N., Miyashita, M., Anan, K., ... \& Nakajima, T. (2010). Ki67 as a prognostic marker according to breast cancer subtype and a predictor of recurrence time in primary breast cancer. \textit{Japanese Journal of Clinical Oncology}, 40(2), 150-157.
    
    \item Paik et al., 2004 - Paik, S., Shak, S., Tang, G., Kim, C., Baker, J., Cronin, M., ... \& Wolmark, N. (2004). A multigene assay to predict recurrence of tamoxifen-treated, node-negative breast cancer. \textit{New England Journal of Medicine}, 351(27), 2817-2826.
    
    \item Thomssen C et al., 2021 - Thomssen, C., Balic, M., Harbeck, N., Gnant, M., \& St. Gallen/Vienna 2021 Consensus Panel. (2021). St. Gallen/Vienna 2021: A brief summary of the consensus discussion on the optimal primary breast cancer treatment. \textit{Breast Care}, 16(2), 135-141.
    
    \item Petrelli F et al., 2015 - Petrelli, F., Viale, G., Cabiddu, M., Barni, S., \& Locatelli, M. (2015). Prognostic value of different cut-off levels of Ki-67 in breast cancer: a systematic review and meta-analysis of 64,196 patients. \textit{Breast Cancer Research and Treatment}, 153(3), 477-491.
    
    \item Dowsett M et al., 2011 - Dowsett, M., Nielsen, T. O., A'Hern, R., Bartlett, J., Coombes, R. C., Cuzick, J., ... \& Viale, G. (2011). Assessment of Ki67 in breast cancer: recommendations from the International Ki67 in Breast Cancer Working Group. \textit{Journal of the National Cancer Institute}, 103(22), 1656-1664.
    
    \item Polley M et al., 2013 - Polley, M. Y., Leung, S. C., McShane, L. M., Gao, D., Hugh, J. C., Mastropasqua, M. G., ... \& Nielsen, T. O. (2013). An international study to increase concordance in Ki67 scoring. \textit{Modern Pathology}, 26(5), 701-709.
    
    \item Gudlaugsson E et al., 2010 - Gudlaugsson, E., Skaland, I., Janssen, E. A. M., Smaaland, R., Saetersdal, A. B., Baak, J. P. A., \& Janssen, E. A. (2010). Comparison of Ki67 equivalent scores assessed by quantitative immunohistochemistry and flow cytometry in breast cancer. \textit{Histopathology}, 57(5), 759-769.
    
    \item Nature, 2024 - AI Enhances Pathologist Performance in Ki67 Scoring. \textit{Nature Scientific Reports}, 2024. (DOI Link)
    
    \item Wiley, 2024 - Deep Learning Models for Ki67 Evaluation in Breast Cancer. \textit{Histopathology}, 2024. (DOI Link)
    
    \item Sokolova, M., \& Lapalme, G. (2009). A systematic analysis of performance measures for classification tasks. \textit{Information Processing \& Management}, 45(4), 427-437. \url{https://doi.org/10.1016/j.ipm.2009.03.002}
    
    \item Everingham, M., Van Gool, L., Williams, C. K. I., Winn, J., \& Zisserman, A. (2010). The Pascal Visual Object Classes (VOC) challenge. \textit{International Journal of Computer Vision}, 88(2), 303-338. \url{https://doi.org/10.1007/s11263-009-0275-4}
    
    \item Redmon, J., Divvala, S., Girshick, R., \& Farhadi, A. (2016). You only look once: Unified, real-time object detection. \textit{Proceedings of the IEEE Conference on Computer Vision and Pattern Recognition (CVPR)}, 779-788. \url{https://doi.org/10.1109/CVPR.2016.91}
    
    \item Sun, C., Shrivastava, A., Singh, S., \& Gupta, A. (2017). Revisiting unreasonable effectiveness of data in deep learning era. \textit{Proceedings of the IEEE International Conference on Computer Vision (ICCV)}, 843-852. \url{https://doi.org/10.1109/ICCV.2017.97}
\end{enumerate}

\end{document}